\documentclass{llncs}

\pdfoutput=1
\input{quipper_demo.sty}
\usepackage[bookmarksopen=true]{hyperref}

\newcommand{\disclaimer}[1]{\let\oldthefootnote\thefootnote%
\def\thefootnote{}\footnotetext[1]{\scriptsize #1}%
\let\thefootnote\oldthefootnote}

\newcommand{\mydisclaimers}{\disclaimer{$^{\ddagger}$This research was supported by NSERC. $^{\dagger}$This research was supported by
the Intelligence Advanced Research Projects Activity (IARPA) via
Department of Interior National Business Center contract number
D12PC00527. The U.S. Government is authorized to reproduce and
distribute reprints for Governmental purposes notwithstanding any
copyright annotation thereon. Disclaimer: The views and conclusions
contained herein are those of the authors and should not be
interpreted as necessarily representing the official policies or
endorsements, either expressed or implied, of IARPA, DoI/NBC, or the
U.S. Government.}}

\begin{document}

\mainmatter 

\title{An Introduction to\\ Quantum Programming in Quipper}
\titlerunning{An Introduction to Quantum Programming in Quipper}

\author{Alexander S. Green\inst{1}$^{\dagger}$ \and Peter LeFanu Lumsdaine\inst{2}$^{\dagger\ddagger}$ \and Neil
  J. Ross\inst{1}$^{\dagger\ddagger}$ \and Peter~Selinger\inst{1}$^{\dagger\ddagger}$ \and Beno\^{i}t Valiron\inst{3}$^{\dagger}$}

\institute{Dalhousie University, Halifax, NS, Canada \\
 \email{agreen@mathstat.dal.ca}, \email{Neil.JR.Ross@Dal.Ca},
 \email{selinger@mathstat.dal.ca} 
 \and Institute of Advanced Studies, Princeton, NJ, U.S.A. \\
 \email{p.l.lumsdaine@gmail.com}
 \and University of Pennsylvania, Philadelphia, PA, U.S.A. \\
 \email{benoit.valiron@monoidal.net}}
 
\authorrunning{A.S. Green, P.L. Lumsdaine, N.J. Ross, P. Selinger, and
  B. Valiron}

\maketitle

\mydisclaimers

\begin{abstract}
Quipper is a recently developed programming language for expressing
quantum computations. This paper gives a brief tutorial introduction
to the language, through a demonstration of how to make use of some of
its key features. We illustrate many of Quipper's language
features by developing a few well known examples of Quantum
computation, including quantum teleportation, the quantum Fourier
transform, and a quantum circuit for addition. 
\end{abstract}

\section{Introduction}
\subsection{Overview}
Quipper {\cite{quipper}} is an embedded functional programming
language for quantum computation. It has been developed as part of
IARPA's QCS project {\cite{BAA}}. The stated goal of the QCS project
is to ``\emph{accurately estimate and reduce the computational
  resources required to implement quantum algorithms on a realistic
  quantum computer}'', with an emphasis on using techniques that have
been developed in the realms of computer science.

In this paper, we will look at how Quipper can be used to implement
existing quantum algorithms, through a close look at some of the
language features that have been added specifically for this task.
Quipper's development was guided by the goal of implementing seven
non-trivial quantum algorithms from the literature
{\cite{BF,BWT,CN,LS,TF,SV,GSE}}. These algorithms were chosen by the
QCS project, and provided to us in modified form.  They cover a broad
spectrum of techniques used in quantum computation. Each algorithm
introduced its own challenges that helped guide the language features
that are now available in Quipper.

We will use simple examples to try to demonstrate the use of Quipper,
and to give insights into the types of problems that the various
language features are useful for.  We will consider three main
stand-alone examples:
\begin{itemize}
\item Quantum teleportation will guide us through: Quipper's
  underlying circuit model, Quipper's primitive operations, quantum
  data-types, generic functions, comments, and labels.
\item The quantum Fourier transform and quantum addition will help us
  look at: recursion, circuit-level operators, boxed circuits, and
  simulation.
\item We will end by looking at Quipper's features that can be used
  to implement quantum oracles,
  including: automatic generation of circuits from classical code,
  synthesis of reversible circuits, and circuit transformations.
\end{itemize}
We will also have a brief look at how Quipper can be used to estimate
the computational resources required by the algorithms that have been
implemented.

In another recent paper \cite{quipper}, we have described in more
detail the rationale behind the various design choices that went into
Quipper, including a high-level overview of, and justification for,
its language features. We also gave more background on general issues
affecting quantum programming languages, and on the implementation of
the language itself. By contrast, the aim of this present paper is to
give a tutorial introduction to Quipper from a programmer's
perspective, using examples that have been chosen to guide readers
through some of Quipper's main features.

\subsection{Quipper as an embedded language}

Quipper has been implemented as an embedded language, using Haskell as
the host language. Therefore, Quipper can be seen as a collection of
data types, combinators, and a library of functions within Haskell,
together with an {\em idiom}, i.e., a preferred style of writing
embedded programs. In this paper, we present Quipper as if it were a
language in its own right, i.e., without presupposing any knowledge of
Haskell.

While the embedded language approach has many advantages (see
{\cite[Sec.~1.3]{Claessen-2001}} for a general discussion), there are
also certain potential pitfalls that programmers should be aware of.
One of these is the temptation to ``escape to the host language'',
i.e., to write general Haskell programs rather than following
Quipper's intended idiom. This can break intended abstractions, and
make the programs less portable in case of implementation changes.
Another drawback of the embedded language approach is that compilation
errors are often difficult to decipher, because the compiler presents
them in terms of concepts of the host language, rather than the
embedded language. Finally, while Haskell is a good fit for Quipper in
many respects, it does lack two features that would be useful for a
quantum programming language: {\em linear types} and {\em dependent
  types}. We must therefore live with checking certain well-formedness
properties of programs at run-time, although they could in principle
be checked by the type-checker in a dedicated language.

\subsection{Quipper's underlying circuit model}

Quipper uses an extended circuit model of quantum
computation. We allow for both quantum and classical wires and operations
within a circuit.  Quantum operations can be controlled by
a classical wire, but not vice versa. A quantum wire can be
explicitly measured, thus creating a classical wire. Quipper's circuit
model also incorporates explicitly scoped ancilla wires, allowing for
an ancilla to only come into scope for the part of the circuit in
which it is used. This is achieved by allowing explicit qubit
initialization and termination within a circuit. 

Using a circuit model leads to three distinct phases of execution:
compile time, circuit generation time, and circuit execution
time. This, in turn, gives rise to an extra distinction among inputs.
Inputs whose value is known at circuit generation time will be called
{\em parameters}; whereas inputs whose value is only known at circuit
execution time will be called {\em inputs}. To keep this distinction
explicit, Quipper introduces three basic types for bits and qubits. We
use the type $\Bool$ for a boolean parameter that is known at circuit
generation time, the type $\Bit$ for a classical boolean input to a
circuit, and the type $\Qubit$ for a quantum input to a
circuit.  A parameter of type $\Bool$ can easily be
converted to an input of type $\Bit$, but not vice versa. Also, because
measurements can only occur at circuit runtime, the outcome of a
measurement is a $\Bit$, not a $\Bool$.

\section{Quipper by example}

\subsection{Quantum teleportation}

\subsubsection{Quipper's primitive operations.}

Although Quipper can be regarded as a language for describing quantum
circuits, when actually developing computations within Quipper it is
often preferable to think in terms of gates being applied in real time
to qubits (or bits) that are held in variables. This procedural
paradigm is the foundation for developing quantum computations in
Quipper, on top of which more powerful higher-order operators are built.

Computations in Quipper take the form of functions. The following
example shows how we can write a simple quantum function in
Quipper.

\begin{code}
\begin{verbatim}
plus_minus :: Bool -> Circ Qubit
plus_minus b = do
    q <- qinit b
    r <- hadamard q
    return r
\end{verbatim}
\end{code}

\noindent 
The first line corresponds to the {\em type} of the
function. We see that the input to the function is a boolean
parameter. The output type of the function is $\Circ~\Qubit$. The
$\Circ$ part of the type is actually a type operator, and is used
to state that the function being defined can have a physical side
effect when it is evaluated (Haskell programmers will recognize this
as a {\em monad}). The $\Qubit$ part of the output type tells us
that the function returns a qubit.  The body of the function usually
starts with the keyword \verb!do!, followed by a block of quantum
operations to be evaluated in the given order. The body of the
\verb!plus_minus! function uses three operations. The \verb!qinit!
operator initializes a new qubit, in the state corresponding to
$b$. Here, $\False$ corresponds to $\ket{0}$ and $\True$
corresponds to $\ket{1}$. The notation tells us that this newly
created qubit is stored in the variable $q$. The operator
\verb!hadamard! applies the Hadamard gate to the qubit $q$,
storing the updated qubit in the variable $r$. The last line
returns the qubit $r$ as the output of the whole function. In
summary, this function introduces a newly initialized qubit in either
of the states $\ket{+}$ or $\ket{-}$ depending on a boolean
parameter. We also note that variables in the function body are used
{\em linearly}: each qubit is written exactly once and read exactly
once. This restriction is imposed by the laws of quantum physics. In
Quipper's syntax, however, it would have been permitted to use the
same name for the two variables $q$ and $r$, and we will
often do so in future examples.

\subsubsection{Circuit generation.}

After defining a quantum function in Quipper, there are various things
we can do with it. The most basic of these is to evaluate the function
to generate a circuit.  When Quipper evaluates a circuit producing
function, the circuit is produced lazily, on-the-fly. This is useful
for defining very large circuits, whereby the whole circuit doesn't
need to be stored in memory. Moreover, circuits can also be
\emph{consumed} lazily, for example by a transformation (see
p.~\pageref{page-transformation}), or by passing instructions
sequentially to an (actual or simulated) quantum computer (see
p.~\pageref{page-simulator}).

A useful operation provided by Quipper is a circuit printing function
that enables the circuits produced by Quipper to be exported in
various formats. For example, to produce a PDF document from the
circuit defined by the above \verb!plus_minus! function, we can use
the built-in Quipper operator \verb!print_simple!. Note that
parameters, but not inputs, must be specified at circuit generation
time. Here, we set the parameter $b$ to $\False$. 

\begin{splitcode}
\begin{verbatim}
print_plus_minus :: IO () 
print_plus_minus = print_simple PDF (plus_minus False)
\end{verbatim} 
  \split \hspace{.7in}\includegraphics[scale=1.5]{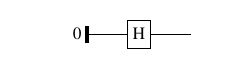}
\end{splitcode}

\noindent 
The circuit diagrams used throughout the rest of this paper
have been created directly from the given code examples. The next
example illustrates how to control a quantum gate. 
This function inputs a qubit and returns a pair of qubits. The
\verb!qnot! operation applies a not-gate to the qubit
$b$. Moreover, the infix operator \verb!`controlled`! causes this
operation to be controlled by the qubit $a$. The overall effect
of the function \verb!share! is to take a qubit in the state
$\alpha\ket{0} + \beta\ket{1}$ and entangle it with a newly
initialized qubit to create the state $\alpha\ket{00} +
\beta\ket{11}$.

\begin{splitcode}
\begin{verbatim}
share :: Qubit -> Circ (Qubit, Qubit) 
share a = do 
  b <- qinit False 
  b <- qnot b `controlled` a 
  return (a,b) 
\end{verbatim} 
\split
  \includegraphics[scale=1.5]{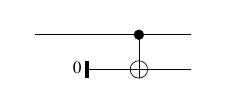}
\end{splitcode}

Previously defined quantum functions can be used as building blocks in
other quantum functions. In fact, they can be used in exactly the same
way as Quipper's built-in operators. In the next example, we use our
previously defined functions, \verb!plus_minus!  and \verb!share!, to
produce a pair of qubits in the Bell state 
$\frac{1}{\sqrt{2}}(\ket{00} + \ket{11})$.

\begin{splitcode}
\begin{verbatim} 
bell00 :: Circ (Qubit, Qubit) 
bell00 = do 
  a <- plus_minus False 
  (a,b) <- share a 
  return (a,b)
\end{verbatim}
\split
  \includegraphics[scale=1.5]{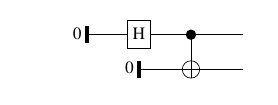}
\end{splitcode}

\subsubsection{A teleportation circuit.}

Let us now consider quantum teleportation (see {\cite{Nielsen-Chuang-2002}}
for an introduction). This involves two parties Alice and Bob. Alice's
goal is to teleport a qubit $q$ to Bob. Alice and Bob must each have
access to a single qubit from an entangled Bell pair $(a,b)$, which we
can produce with the above \verb!bell00! function. We can think of
Alice's role in terms of a function that inputs the two qubits $q$ and
$a$. The output of the function will be a pair of classical bits,
produced by Alice by applying some unitary gates and then measuring
both qubits.

\begin{splitcode}
\begin{verbatim}
alice :: Qubit -> Qubit -> Circ (Bit,Bit)
alice q a = do
    a <- qnot a `controlled` q
    q <- hadamard q
    (x,y) <- measure (q,a)
    return (x,y)
\end{verbatim}
\split
\includegraphics[scale=1.5]{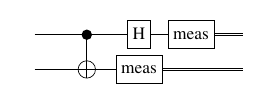}
\end{splitcode}

\noindent
Note that the function \verb!measure! has been applied to a pair of
qubits. In Quipper's syntax, this is simply an abbreviation for
measuring both qubits in the pair. This abbreviated syntax is possible
because the Quipper operator \verb!measure! is a {\em generic}
operator: it can be applied to any data structure containing qubits,
and returns a corresponding data structure containing bits. Another 
example of a generic Quipper operator is \verb!cdiscard!, which can be
applied to any data structure
containing classical bits. It is used in Bob's part of the
teleportation protocol:

\begin{splitcode}
\begin{verbatim}
bob :: Qubit -> (Bit,Bit) -> Circ Qubit
bob b (x,y) = do
    b <- gate_X b `controlled` y
    b <- gate_Z b `controlled` x
    cdiscard (x,y)
    return b
\end{verbatim}
\split
\includegraphics[scale=1.5]{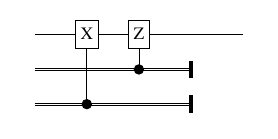}
\end{splitcode}

\noindent
The following function ties all the pieces of the teleportation
example together. We can see that a Bell state is created, which is
then used by Alice, along with the input qubit, to create a pair of
classical bits. These are passed to Bob along with his qubit from the
Bell state. The generated circuit diagram shows that Quipper joined
together the various steps
as expected.

\begin{splitcode}
\begin{verbatim}
teleport :: Qubit -> Circ Qubit
teleport q = do
    (a,b) <- bell00
    (x,y) <- alice q a
    b <- bob b (x,y)
    return b
\end{verbatim}
\split
\hspace{-1in}
\includegraphics[scale=1.3]{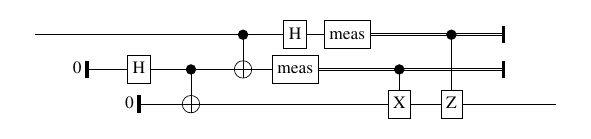}
\end{splitcode}

\subsubsection{Quantum data types and generic functions.}  

{\em Quantum data types} are types that are built up from $\Qubit$
by means of data constructors, such as tuples and lists. For example,
$\verb!(Qubit,[Qubit])!$ is the type whose elements are pairs of a
qubit and a (variable but finite length) list of qubits. Every quantum
data type, such as $qa = \verb!(Qubit,[Qubit])!$, has an associated
{\em classical data type}, such as $ca = \verb!(Bit,[Bit])!$, and {\em
  boolean data type}, such as $ba = \verb!(Bool,[Bool])!$. We say that
$qa$, $ca$, and $ba$ have the same {\em shape}, but different {\em
  leaf types}. A Quipper function is called {\em generic} if it can
act on data types of any shape. 

We have already seen several examples of generic built-in Quipper
functions, namely \verb!measure!, \verb!cdiscard!, and
\verb!print_simple!. However, what makes generic functions
particularly useful in Quipper is the fact that it is easy to create
new {\em user-defined} generic functions. We will now illustrate this
feature by defining a generic version of the teleportation circuit. 

In Quipper, the keyword \verb!QShape! is used to declare that three
types $qa$, $ca$, and $ba$ are the quantum, classical, and boolean
version of some data type. To define a generic version of the
\verb!plus_minus! function, we replace $\Bool$ and $\Qubit$ in
its type by such a pair of related $ba$ and $qa$:

\begin{code}
\begin{verbatim}
plus_minus_generic :: (QShape ba qa ca) => ba -> Circ qa
plus_minus_generic a = do 
    qs <- qinit a
    qs <- mapUnary hadamard qs
    return qs
\end{verbatim}
\end{code}

\noindent
We note that the \verb!qinit! function is already generic. The
operator \verb!mapUnary! maps a function of type
$\Qubit\to\Circ~\Qubit$ over every qubit in a quantum data
structure.  To extend the \verb!share! function, we use the function
\verb!qc_false! which generates a boolean data structure of the
correct shape, with every boolean set to $\False$. The
\verb!mapBinary! function is similar to \verb!mapUnary!, but maps a
function of the type $\Qubit\to\Qubit\to\Circ\,(\Qubit,\Qubit)$ over
every corresponding pair of qubits from two quantum data structures of
the same shape. We also use the built-in \verb!controlled_not!
operation.

\begin{code}
\begin{verbatim}
share_generic :: (QShape a qa ca) => qa -> Circ (qa, qa)
share_generic qa = do 
    qb <- qinit (qc_false qa)
    (qb, qa) <- mapBinary controlled_not qb qa
    return (qa, qb)
\end{verbatim}
\end{code}

Updating the \verb!bell00! function requires a little more thought, as
we now need to explicitly know the shape of the data being teleported
in order to generate enough Bell pairs. This is achieved by
adding a shape argument to the function, which can then be used by the
call to \verb!plus_minus_generic!.

\begin{code}
\begin{verbatim}
bell00_generic :: (QShape a qa ca) => a -> Circ (qa, qa)
bell00_generic shape = do 
    qa <- plus_minus_generic shape
    (qa, qb) <- share_generic qa
    return (qa, qb)
\end{verbatim}
\end{code}

\pagebreak
\noindent
The changes to Alice's function are very similar to those we have seen
already.

\begin{code}
\begin{verbatim}
alice_generic :: (QShape a qa ca) => qa -> qa -> Circ (ca,ca)
alice_generic q a = do 
    (a, q) <- mapBinary controlled_not a q
    q <- mapUnary hadamard q
    (x,y) <- measure (q,a)
    return (x,y)
\end{verbatim}
\end{code}

For Bob's function, we need a way of mapping classically controlled
$X$- and $Z$-rotations over the input bits and qubits. The function
\verb!mapBinary_c! is similar to \verb!mapBinary!, except that it
expects a function of type 
$\Qubit\to\Bit\to\Circ\,(\Qubit,\Bit)$. Also, whereas the
\verb!controlled_not! function is a built-in operator, the classically
controlled $X$ and $Z$ rotations are not. We use a \verb!where!
clause to define a generic \verb!controlled_gate! function
locally.

\begin{code}
\begin{verbatim}
bob_generic :: (QShape a qa ca) => qa -> (ca,ca) -> Circ qa
bob_generic b (x,y) = do
     (b, y) <- mapBinary_c (controlled_gate gate_X) b y
     (b, x) <- mapBinary_c (controlled_gate gate_Z) b x
     cdiscard (x,y)
     return b
  where
    controlled_gate gate b x = do
      gate b `controlled` x
      return (b,x)
\end{verbatim}
\end{code}

\noindent 
The various parts of the generic teleportation function can now be
tied together.

\begin{code}
\begin{verbatim}    
teleport_generic :: (QData qa) => qa -> Circ qa
teleport_generic q = do
    (a,b) <- bell00_generic (qc_false q)
    (x,y) <- alice_generic q a
    b <- bob_generic b (x,y)
    return b
\end{verbatim}
\end{code}

\noindent
Note that a generic Quipper function defines a {\em family} of
circuits, one for each data type. To be able to print specific members
of this family, we must replace the \verb!print_simple! operator by
the more general \verb!print_generic!. The difference is that
\verb!print_generic! takes additional arguments to determine which
instance of the circuit family to print. We show examples for
teleporting a pair of qubits, and a list of three qubits:
\begin{code}
\begin{verbatim}
print_generic PDF teleport_generic (qubit, qubit)
\end{verbatim}
\end{code}
\vspace{-3ex}
\[
\includegraphics[width=.7\linewidth]{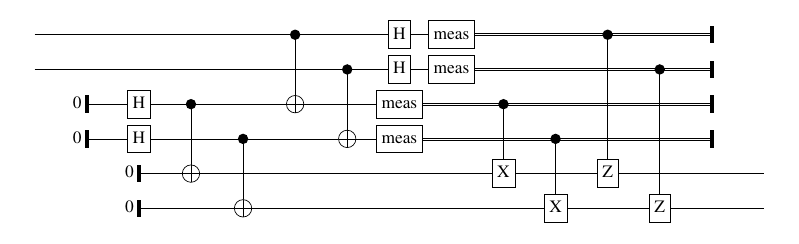}
\]
\pagebreak
\begin{code}
\begin{verbatim}
print_generic PDF teleport_generic [qubit,qubit,qubit]
\end{verbatim}
\end{code}
\vspace{-3ex}
\[
\includegraphics[width=.9\linewidth]{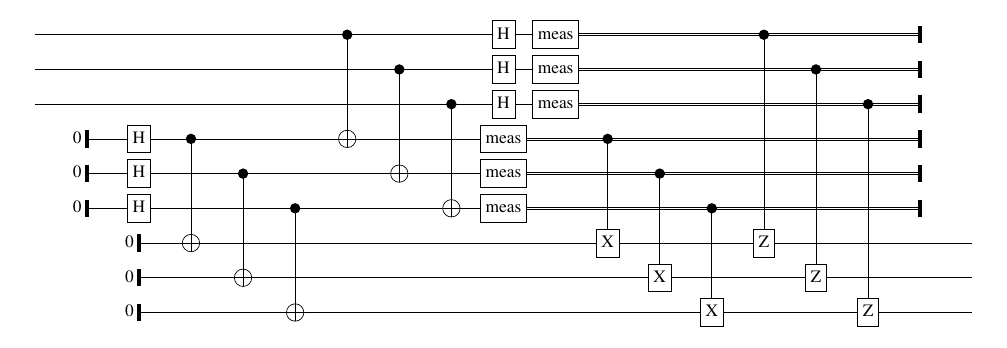}
\]

\subsubsection{Comments and labels.}

When reading very large circuits, it is sometimes hard to keep track
of what each part of the circuit is doing, or which wires certain
variables correspond to. As a convenience to the programmer, Quipper
offers a way of adding comments and labels to a circuit:

\begin{code}
\begin{verbatim}
teleport_generic_labeled :: (QData qa) => qa -> Circ qa
teleport_generic_labeled q = do
    comment_with_label "ENTER: bell00" q "q"
    (a,b) <- bell00_generic (qc_false q)
    comment_with_label "ENTER: alice" (a,b) ("a","b")
    (x,y) <- alice_generic q a
    comment_with_label "ENTER: bob" (x,y) ("x","y")
    b <- bob_generic b (x,y)
    return b
\end{verbatim}
\end{code}
\vspace{-1ex}
\[
\includegraphics[width=\textwidth]{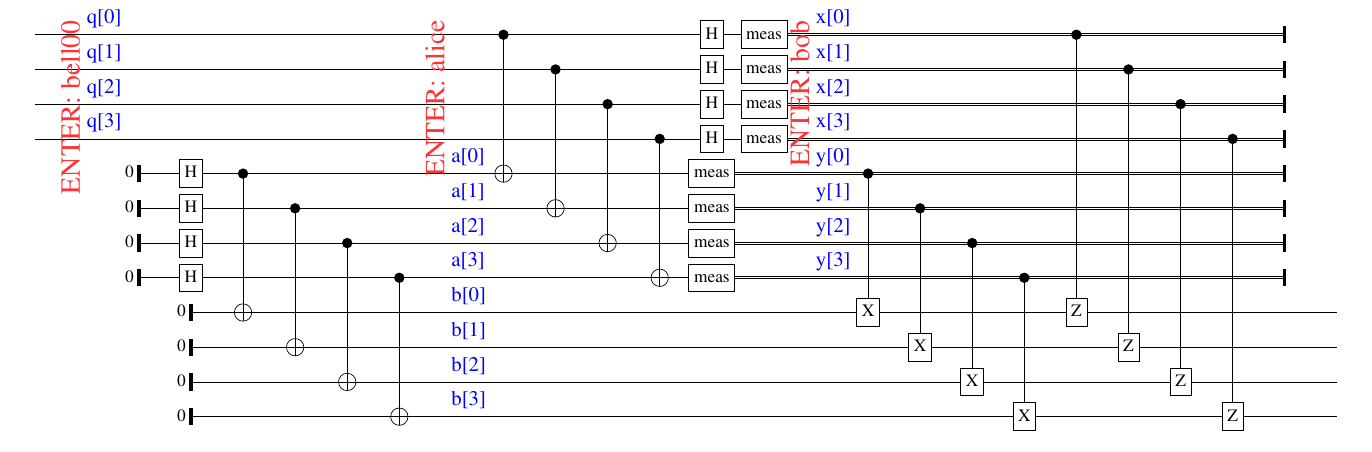}
\]

\subsection{The quantum Fourier transform and quantum addition}

\subsubsection{Recursion.}
In Quipper it is possible to write circuit producing functions that
are recursive over any parameters known at circuit generation
time. Notably, we can write functions that are recursive over the
shape of an input, such as a list of qubits. For example, consider the
quantum Fourier transform, or QFT, which lends itself nicely to a
recursive definition.  The function \verb!qft'! is defined over a list
of qubits. We provide two base cases for the recursion. If the input
list is empty, the circuit itself is empty. If the input is a
singleton qubit, then the QFT is just the Hadamard gate. For the
recursive case, the circuit for the QFT for $n+1$ qubits consists of
the circuit for the $n$ qubit QFT, followed by a set of rotations over
all $n+1$ qubits. This set of rotations can also be defined in terms
of a recursive function, which we call \verb!rotations!. Also,
$\verb!rGate!\,m$ is a built-in Quipper operator that
represents the $z$-rotation by $\frac{2\pi i}{2^{m}}$.

\begin{code}
\begin{verbatim}
qft' :: [Qubit] -> Circ [Qubit]
qft' [] = return []
qft' [x] = do 
  hadamard x
  return [x]
qft' (x:xs) = do 
  xs' <- qft' xs
  xs'' <- rotations x xs' (length xs')
  x' <- hadamard x
  return (x':xs'')
 where
   rotations :: Qubit -> [Qubit] -> Int -> Circ [Qubit]
   rotations _ [] _ = return []
   rotations c (q:qs) n = do 
     qs' <- rotations c qs n
     let m = ((n + 1) - length qs)
     q' <- rGate m q `controlled` c
     return (q':qs')
\end{verbatim}
\end{code}

\vspace{1ex}
\noindent
The function \verb!qft'! expects its list of input qubits in
little-endian order, but returns the output in big-endian
order. Because this is confusing, we wrap it in another function
\verb!qft_big_endian!, which simply reverses the order of the input
qubits.  In Quipper, this is done not by swapping wires in a circuit,
but by reordering {\em references} to wires; Quipper will attach the
rest of the circuit appropriately.

\begin{code}
\begin{verbatim}
qft_big_endian :: [Qubit] -> Circ [Qubit]  
qft_big_endian qs = do
  comment_with_label "ENTER: qft_big_endian" qs "qs"
  qs <- qft' (reverse qs)
  comment_with_label "EXIT: qft_big_endian" qs "qs"
  return qs
\end{verbatim}
\end{code}
\vspace{-2ex}
\[
\includegraphics[width=.8\linewidth]{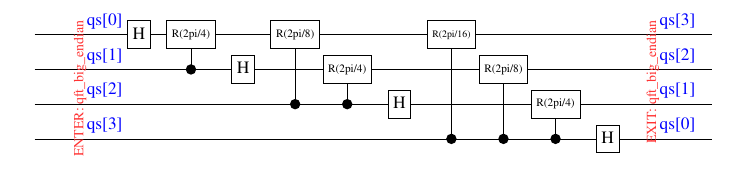}
\]

\subsubsection{Circuit-level operations.}

Most operators we have seen so far work at the level of gates, i.e.,
their effect is to append gates one by one to a circuit under
construction. Quipper also has the idiom of circuit-level operations,
which are operations that can be applied to circuits as a whole. One
example is the printing of circuits, but there are also circuit-level
operations that can be used while constructing circuits. These often
take a circuit generating function as input, and produce a new circuit
generating function as an output, which can then be used just like any
other circuit generating function. A useful example is the operator
\verb!reverse_generic_endo!, which reverses a whole circuit. The
following function computes the inverse of the QFT.

\begin{code}
\begin{verbatim}
inverse_qft_big_endian :: [Qubit] -> Circ [Qubit]
inverse_qft_big_endian = reverse_generic_endo qft_big_endian
\end{verbatim}
\end{code}
\vspace{-2ex}
\[
\includegraphics[width=.8\linewidth]{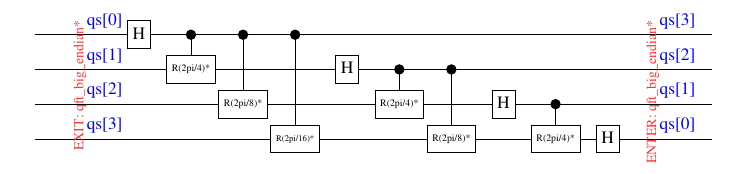}
\]

\subsubsection{A quantum adder.}

As an application of the QFT, we look at a quantum circuit that
performs addition {\cite{draper}}, without the use of ancilla
qubits. The circuit uses a QFT as a basis change. The inverse QFT is
then applied at the end to change back to the computational
basis. The part of the circuit that
performs the actual addition, between the two uses of the QFT, once again
lends itself to a recursive definition.

\begin{code}
\begin{verbatim}
qft_adder :: [Qubit] -> [Qubit] -> Circ ()
qft_adder _ [] = return ()
qft_adder as (b:bs) = do
  qft_adder' as b 1
  qft_adder (tail as) bs
 where
   qft_adder' :: [Qubit] -> Qubit -> Int -> Circ [Qubit]
   qft_adder' [] _ _ = return []
   qft_adder' (a:as) b n = do
     b <- rGate n b `controlled` a
     qft_adder' as b (n+1)
\end{verbatim}
\end{code}

\vspace{1ex}
The pattern of applying an initial computation, followed by some
operation, followed by the inverse of the initial computation, is
quite common in quantum computation. For this reason, Quipper provides
a circuit-level operator \verb!with_computed!, which automatically
takes care of applying the inverse computation at the end. We use this
here to complete the quantum addition circuit, using the QFT as the
initial computation to be inverted at the end.

\begin{code}
\begin{verbatim}
qft_add_in_place :: [Qubit] -> [Qubit] -> Circ ([Qubit], [Qubit])
qft_add_in_place a b = do
  label (a,b) ("a","b")
  with_computed (qft_big_endian b) $ \b' -> do
    qft_adder a (reverse b')
  label (a,b) ("a","b")
  return (a,b)
\end{verbatim}
\end{code}
\vspace{-2ex}
\[
\hspace{-2ex}\includegraphics[width=1.05\linewidth]{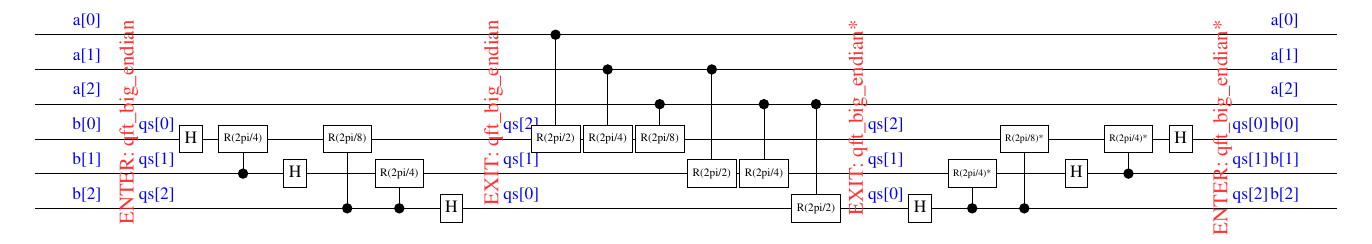}
\]

\subsubsection{Boxed subcircuits.}

In many quantum algorithms, the same subcircuit is reused multiple
times, which can cause a lot of duplication in circuits. Quipper helps
alleviate such duplication by providing a hierarchical model of
circuits, in the form of boxed subcircuits. A circuit can be
\emph{boxed}, and then reused multiple times as a subcircuit in a
larger circuit. This means that the boxed subcircuit only needs to be
generated once, and then a call to the boxed subcircuit is placed in
the main circuit, whenever the subcircuit would appear. Quipper also
permits an iteration count to be attached to a boxed subcircuit call. 

A subcircuit can be boxed by using the \verb!box! operator, which
takes as its arguments a name and a function to be boxed. Here, we
replicate the previous example, but with the QFT boxed.

\begin{code}
\begin{verbatim}
qft_add_in_place_boxed :: [Qubit] -> [Qubit] -> Circ ([Qubit], [Qubit])
qft_add_in_place_boxed a b = do
  label (a,b) ("a","b")
  with_computed (box "QFT" qft_big_endian b) $ \b' -> do
   qft_adder a (reverse b')
  label (a,b) ("a","b")
  return (a,b)
\end{verbatim}
\end{code}
\vspace{-2ex}
\[
\includegraphics[width=0.9\linewidth,page=1]{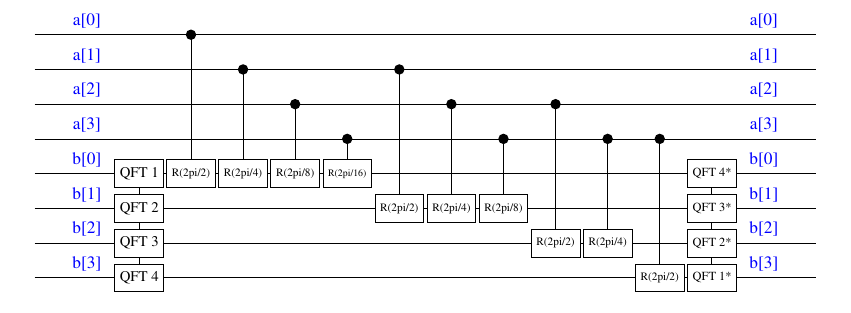}
\]
\[
\includegraphics[width=0.8\linewidth,page=2]{qft_add_boxed}
\]

\subsubsection{Simulation of circuits.}\label{page-simulator}

Unlike many quantum programming languages in the literature, Quipper
was not designed as a front-end language for a quantum simulator;
rather, it was designed to control an actual (future) quantum
computer. Therefore, non-physical operations are not provided in
Quipper. Nevertheless, during development and testing (and in the
absence of an actual quantum computer), it is useful to be able to run
simulations. Quipper provides three different simulators, which can be
used depending on which gates are used within a circuit.
\begin{itemize}
\item Classical simulation - efficiently simulates classical circuits.
\item Stabilizer simulation - efficiently simulates Clifford group circuits {\cite{stabilizer}}.
\item Quantum simulation - simulates any circuit (with exponential overhead).
\end{itemize}
The simulators are generic: they take any circuit producing function and
convert it into a function acting on the boolean counterparts to the
quantum data types used in the circuit. Both the stabilizer
simulator, and the quantum simulator are probabilistic. 

\subsection{Quantum circuits from classical functions}

\subsubsection{Generating circuits from classical code.}

A notable feature of Quipper is the ability to automatically generate
reversible circuits from ordinary functional programs. This is
achieved by inserting the Quipper keyword \verb!build_circuit! right
before the classical function definition. This causes Quipper to
define a new circuit generating function, with the same name as the
given classical function, preceded by \verb!template_!, where any
$\Bool$ arguments in the type are changed to $\Qubit$. We found that
this language feature is useful when defining many of the oracles that
appear in quantum algorithms, as they are often of a classical nature,
but need to be applied to a quantum register. We have used this
feature, for example, to implement a quantum library for real
fixed-point arithmetic. The following example shows a single-bit full
adder. A quantum function named \verb!template_adder! will be
automatically generated.

\begin{code}
\begin{verbatim}
build_circuit
adder :: (Bool,Bool,Bool) -> (Bool,Bool)
adder (a,b,carry_in) = (s,carry_out)
 where
  s = bool_xor (bool_xor a b) carry_in
  carry_out = (a && b) || (a && carry_in) || (b && carry_in)
\end{verbatim}
\end{code}

\noindent
The helper function \verb!unpack! is used to tidy up the type of any
circuit produced using the \verb!build_circuit! keyword, by removing
some unnecessary occurrences of the $\Circ$ operator.

\vspace{-2ex}
\begin{splitcode}
\vspace{1ex}
\begin{verbatim}
adder_circ :: (Qubit,Qubit,Qubit) 
    -> Circ (Qubit,Qubit)
adder_circ = unpack template_adder
\end{verbatim}
\split
\includegraphics[width=\linewidth]{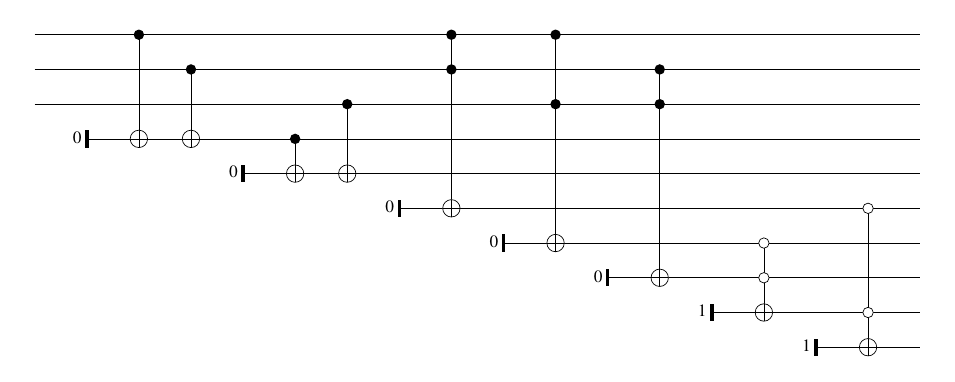}
\end{splitcode}

The \verb!build_circuit! feature is implemented using a Haskell
extension known as {\em Template Haskell\/}; this gives programs
access to their own syntax tree in parsed form. Because of this
generality, essentially arbitrary Haskell functions can be used with
the \verb!build_circuit! keyword. However, the programmer must supply
quantum templates for any library functions that are used, unless they
are among the standard templates already predefined by Quipper.

\vspace{-1ex}

\subsubsection{Synthesis of reversible circuits.}

The circuit produced by \verb!adder_circ! is not a self-contained
reversible circuit, as the automatic transformation introduces ancilla
qubits that may be left in an indeterminate state, possibly entangled
with the outputs. The Quipper operator \verb!classical_to_reversible!
turns a circuit $f :: a \to \Circ~b$ into a reversible circuit $f' ::
(a,b) \to \Circ\,(a,b)$, ensuring that any ancillas are suitably
un-computed and terminated, provided that $f$ uses only reversible
primitives.

\vspace{1ex}
\begin{code}
\begin{verbatim}
adder_reversible :: ((Qubit,Qubit,Qubit),(Qubit,Qubit)) 
   -> Circ ((Qubit,Qubit,Qubit),(Qubit,Qubit))
adder_reversible = classical_to_reversible adder_circ
\end{verbatim}
\end{code}
\vspace{-2ex}
\[
\includegraphics[width=\linewidth]{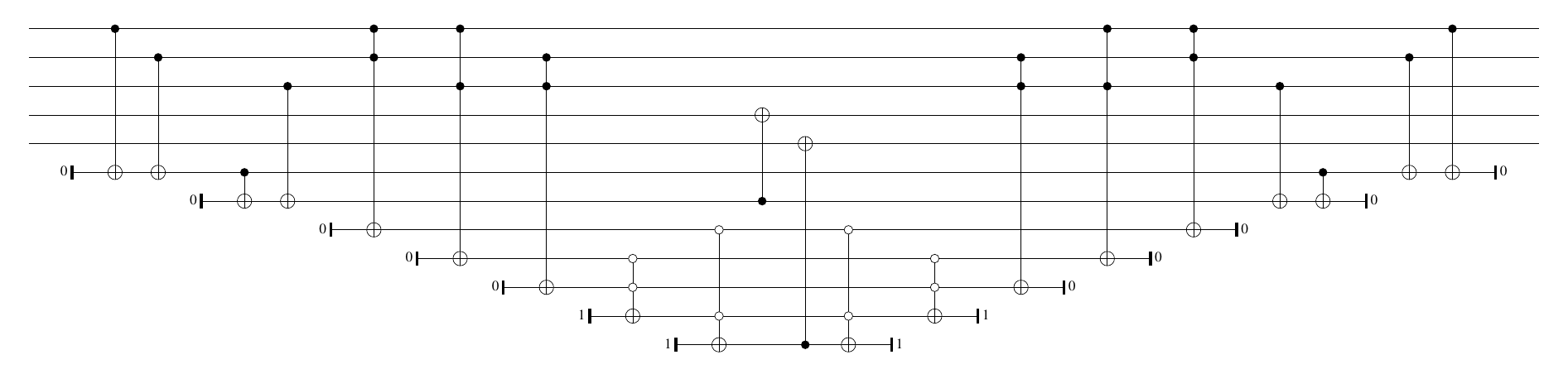}
\]

\subsubsection{Circuit transformations.}\label{page-transformation}

Quipper provides a means for transforming circuits, on-the-fly, at
circuit generation time. This allows for transformations such as gate
decompositions, or adding certain types of error-correcting
codes. Quipper provides some pre-defined transformers, as well as an
extensible framework for user-defined transformers. Example
transformers include the simulators, as well as a transformer to
decompose circuits to only binary gates, or binary gates plus the
Toffoli gate. In the following example, we apply the binary gate
decomposition transformer to the adder circuit.

\vspace{1ex}
\begin{code}
\begin{verbatim}
adder_circ_b :: (Qubit,Qubit,Qubit) -> Circ (Qubit,Qubit)
adder_circ_b = decompose_generic Binary adder_circ
\end{verbatim}
\end{code}
\vspace{-2ex}
\[
\includegraphics[width=\linewidth]{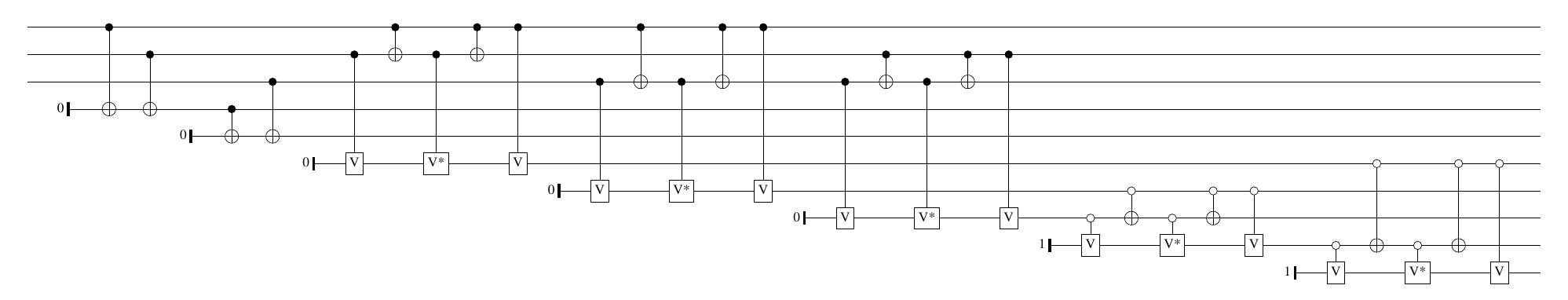}
\]

\section{Final remarks}

\subsection{Scalability and resource estimation}

As we have seen, there are various things that Quipper can do with a
generated circuit. However, when defining large circuits, it isn't
always feasible to generate the circuit in its entirety. Quipper
provides a mechanism by which one can count the resources associated
with a circuit (e.g., number of gates, number of qubits, number of
ancillas). Combining this feature with boxed subcircuits, we have been
able to do resource estimation for some very large circuits. For
example, our Quipper implementation of the triangle finding algorithm
{\cite{TF}} produces a circuit containing over 30 trillion gates,
which can be counted in under two minutes on a 1.2GHz
laptop.
\vspace{-0.5ex}

\subsection{Prior art}

There have been a number of quantum programming languages introduced
in the literature (see {\cite{GaySJ:quapls}}). Among the languages that have
actually been implemented are \"Omer's QCL {\cite{qcl}}, a C-style
language optimized for quantum simulation; the Quantum IO Monad
{\cite{Altenkirch-Green-2009}}, which is a quantum programming
language also embedded in Haskell; and Giles's LQPL
{\cite{lqpl}}, a functional quantum programming language with
linear types.  However, most of the languages that can be found in the
literature are not shown to be scalable to large problem sizes. 

The problem of generating circuit descriptions from functional
programs has also been studied outside of the realm of quantum
computing; see, e.g., {\cite{lava,Claessen-2001}}.

\vspace{-0.5ex}

\subsection{Conclusion}

Quipper has many language features, and only a selection of them have
been discussed in this introductory paper. The Quipper distribution
also includes some libraries of commonly used quantum functions. For
example, we provide an extensive library of arithmetic functions, both
for integer arithmetic and fixed-point real arithmetic; and functions
for random access to a quantum register using a quantum
index. Although Quipper is still in active development, we feel that
the current stable release is a full-featured and scalable
language. Many of the improvements that we are hoping to make are to
the type system, such as introducing linear types, which will allow
for more type errors to be caught at the initial compilation stage, as
opposed to at circuit generation time. 

\bibliographystyle{splncs03}
\bibliography{quipper_demo}

\end{document}